\documentclass[a4paper,11pt]{article}
\usepackage{epsf,latexsym,amsmath}
\usepackage{times}
\usepackage{amssymb}
\usepackage{graphicx}
\input{epsf}
\begin{document}

\title{An efficient quantum algorithm for colored Jones polynomials}

\begin{center}
{ \Large \bf  An efficient quantum algorithm for\\
colored Jones polynomials}
\end{center}
\vspace{24pt}

\noindent {\large
{\sl Silvano Garnerone}}\\
\noindent Dipartimento di Fisica,
Politecnico di Torino,\\
corso Duca degli Abruzzi 24, 10129 Torino (Italy);\\ 
E-mail: silvano.garnerone@polito.it \\

\noindent {\large
{\sl Annalisa Marzuoli}}\\
\noindent Dipartimento di Fisica Nucleare e Teorica,
Universit\`a degli Studi di Pavia\\
 and INFN, Sezione di Pavia, 
via A. Bassi 6, 27100 Pavia (Italy);\\ 
E-mail: annalisa.marzuoli@pv.infn.it \\

\noindent {\large
{\sl Mario Rasetti}}\\
\noindent Dipartimento di Fisica,
Politecnico di Torino,\\
corso Duca degli Abruzzi 24, 10129 Torino (Italy);\\ 
E-mail: mario.rasetti@polito.it \\

\begin{abstract}
We construct a quantum algorithm to approximate efficiently the colored Jones polynomial of the plat presentation 
of any oriented link $L$ at a fixed root of unity $q$. Our construction is based on $SU(2)$ Chern-Simons topological 
quantum field theory (and associated Wess-Zumino-Witten conformal field theory) and exploits the $q$-deformed spin 
network as computational background.

As proved in (S. Garnerone, A. Marzuoli, M. Rasetti,
quant--ph/ 0601169),
 the colored Jones polynomial can be evaluated in a number 
of elementary steps, bounded from above by a linear function of the number of crossings of the link, and polynomially 
bounded with respect to the number of link strands.  
Here we show  that the Kaul unitary representation of colored oriented braids used
there can be efficiently 
approximated on a standard quantum circuit.
\end{abstract}

\section{Introduction.}
The present paper is the natural completion of previous work 
\cite{GaMaRa1,MaRa1,MaRa2} which focused on the connections between 
quantum information, automata theory and topological invariants. In \cite{GaMaRa1} the problem of the evaluation 
of the Jones invariant of knots, and its colored extension, has been formalized in the framework of quantum 
formal language theory. Crucial ingredient there is the q-deformed spin-network automaton, first introduced in 
\cite{GaMaRa1}, where it was shown that the colored Jones polynomials can be seen as the expectation value for a 
particular evolution of the automaton. This approach to the problem of evaluating link invariants together with 
further considerations appeared in \cite{AJL,WY} are used here to construct a quantum algorithm for the 
approximation of the colored Jones polynomials at $q = \exp\left(\frac{2\pi i}{k+2}\right)$, $k$ 
integer $\geq 1$. In this note we focus on the basic structure of the algorithm. In the following a basic knowledge 
of knot theory and its relation to the braid group will be assumed (see, e.g., \cite{GaMaRa2}). For a discussion of the 
connections between links invariants, topological quantum field theories and the $q$-deformed spin-network simulator 
we refer the reader to \cite{GaMaRa1}.

Two basic steps are needed in order to construct a quantum algorithm for the approximation of link invariants. 
First, one has to define a unitary representation of the braid group, or of its colored extension. Second, the 
function (Markov trace \cite{Jon} or the suitable topological quantum field theory (TQFT) expectation value \cite{Kau}) 
which allows us to obtain the topological invariant from the unitary representation must be shown to be efficiently 
computable. In section 2 we shall focus on the first step, defining the \textit{Kaul representation} for the colored 
braid group \cite{Kau}. In section 3 we shall prove that such representation and the associated invariant admit an 
efficient implementation in terms of a set of universal quantum gates. A standard procedure is used to access the 
information encoded in the final state of the quantum evolution, which was applied for the first time in this 
context in \cite{AJL}.

\section{Colored Jones polynomial\\ and SU(2) Chern-Simons theory.}
In this section we review the unitary representation of the colored braid group provided by R.K.Kaul in \cite{Kau}, 
referring to the paper for a complete and detailed description of the representation.

\begin{figure}
\begin{center}
 \includegraphics[height=3cm]{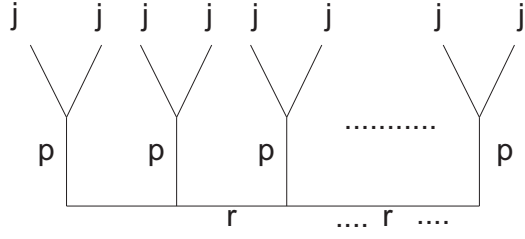}
\end{center}
\caption{\footnotesize{Conformal block associated to the basis vectors in the Kaul representation. To each (j,p,r)-type label 
can be associated different quantum numbers according to the rules of the theory.}}
\label{conformalblock}
\end{figure}

\noindent In \cite{Wit} Witten proved that in a topological quantum field theory link invariants can be obtained 
as expectation values of Wilson loop operators. He also provided a connection between the 3-dimensional 
Chern-Simons TQFT and the 2-dimensional WZW conformal field theory (CFT) living on the boundaries of the 
3-dimensional manifold ${\cal{M}}^{(3)}$. Such connection allows us to build a finite dimensional Hilbert space 
${\mathfrak{H}}$ that can be used to evaluate topological invariants for links embedded in ${\cal{M}}^{(3)}$. 
The basis vectors of ${\mathfrak{H}}$ are associated to conformal blocks of the CFT. Conformal blocks admit a 
graphical realization in terms of unrooted binary trees. Kaul \cite{Kau} provided a unitary representation ${\mathfrak{U}}$ 
of the colored braid group in the Hilbert space spanned by a set of basis vectors corresponding to the conformal 
block of fig(\ref{conformalblock}). In the associated basis the action of ${\mathfrak{U}}$ restricted to the 
odd-indexed braid generators is diagonal, while the action over the even-indexed braid matrices is diagonalized 
only after a change of basis is performed by a duality transformation ${\bf A}$ (corresponds to a $q-3nj$ symbol).

The representation for the generators of the groupoid of colored oriented braids in the basis $|\Phi^{\left( 
\hat{j}_1,\hat{j}_2,...,\hat{j}_{2m} \right) }_{\left(\textbf{p};\textbf{r}\right) }\rangle $, corresponding to 
the conformal block of figure (\ref{conformalblock}), is provided by
\begin{equation}
\left[ B^{2l+1}\right]_{\left( \textbf{p};\textbf{r}\right)}^{\left( \textbf{p'};\textbf{r'}\right)} 
|\Phi^{\left( ...,\hat{j}_{2l+1},\hat{j}_{2l+2},... \right)}_{\left(\textbf{p'};\textbf{r'}\right )}
\rangle = \lambda_{p_{l}}\left( \hat{j}_{2l+1},\hat{j}_{2l+2}\right) \delta^{\textbf{p}}_{\textbf{p'}}
\delta^{\textbf{r}}_{\textbf{r'}} |\Phi^{\left( ...,\hat{j}_{2l+2},\hat{j}_{2l+1},... 
\right)}_{\left(\textbf{p};\textbf{r}\right)} \rangle \; , 
\end{equation} 
and
\begin{equation}
\left[ B^{2l}\right]_{\left( \textbf{p};\textbf{r}\right)}^{\left( \textbf{p'};\textbf{r'}\right)} 
|\Phi^{\left( ...,\hat{j}_{2l},\hat{j}_{2l+1},... \right)}_{\left(\textbf{p'};\textbf{r'}\right)} \rangle 
=\sum_{\left( \textbf{q};\textbf{s}\right) } \lambda_{q_{l}}\left( \hat{j}_{2l},\hat{j}_{2l+1}\right)
A_{\left( \textbf{p};\textbf{r}\right)}^{\left( \textbf{q};\textbf{s}\right)}
A_{\left( \textbf{q};\textbf{s}\right)}^{\left( \textbf{p'};\textbf{r'}\right)}
|\Phi^{\left( ...,\hat{j}_{2l+1},\hat{j}_{2l},... \right)}_{\left(\textbf{p'};\textbf{r'}\right)} \rangle 
\end{equation} 
where all the quantum numbers refer to the irreps of the quantum group $SU(2)_q$. The eigenvalues $\lambda$ 
for right-handed half twists 
depend on the relative orientation of the strands as follows: 
\begin{description}
\item{} if the orientation is the same then $\lambda_t(\hat{j},\hat{j'})=(-)^{j+j'-t}q^{(c_j+c_j')/2+c_{min(j,
j')}-c_t/2}$; 
\item{} if the orientation is different then $\lambda_t(\hat{j},\hat{j'})=(-)^{t-\left|j-j'\right|}q^{-\left | 
c_j-c_j'\right|/2+c_t/2}$,
\end{description}
 where $c_j=j(j+1)$ is the quadratic Casimir operator.
 
\noindent A full specification of the duality matrices $A$ is provided by the following notation
$$
A_{\textbf{m}}^{\textbf{n}}\left[ 
\begin{tabular}{cc}
$j_1$ & $j_2$ \\
 $\vdots$ & $\vdots$ \\
 $j_{2m-1}$ & $j_{2m}$ \\
\end{tabular} \right].
$$ 

\noindent The pictorial realization described in \cite{GaMaRa1} allows us to interpret the multi-labels $\textbf{m}$ and 
$\textbf{n}$ as intermediate angular momenta of a particular binary coupling scheme for the $\left\lbrace 
j_1,...,j_{2m}\right\rbrace$ incoming angular momenta.

\noindent This unitary representation, that we refer to as \textit{Kaul representation}, allows us to evaluate 
the Wilson loop operator for an arbitrary link $L$ presented as the platting of a colored oriented braid $B$. 
The braid is defined as a word written in the generators of the braid group $B_{2m}$ of index $2m$. The link 
invariant is given by
\begin{equation}
V\left( L\right)=\prod_{i=1}^{m}\left[ 2j_{i}+1\right] \langle \Phi_ {\left(\textbf{0};\textbf{0}\right)} 
^{\bf{c'}}\, | W_B \, |\Phi_ {\left(\textbf{0};\textbf{0}\right)} ^{\textbf{c}} \rangle \; , 
\label{expect} 
\end{equation} 
where ${\bf c}$ and ${\bf c'}$ compactly denote the coloring of the link components. The state 
$|\Phi_ {\left(\textbf{0};\textbf{0}\right)}^{\textbf{c}}\rangle$, in which the expectation value is evaluated, 
has all the internal quantum numbers equal to 0. Formula (\ref{expect}) is a theorem, proved in \cite{Kau}. The link 
invariant defined in this way is the colored Jones polynomial. 

\section{Qubit model for the Kaul representation.} 
\begin{figure}
\begin{center}
\includegraphics[height=5cm]{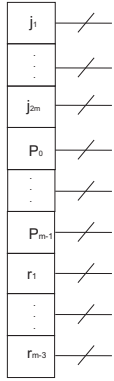}
\caption{\footnotesize{Qubit register for the representation space.}}
\end{center}
\label{register}
\end{figure}

In this section we show how to implement with a quantum circuit the Kaul unitary representation of the braid group. 
We first describe a procedure to encode the basis states of the representation into a register of $n$ qubits, then 
we show how to realize efficiently in such representation the images of the braid group generators.  
\noindent Basis states of our quantum system are labelled by the quantum numbers on the strands of the link and 
by internal quantum numbers corresponding to intermediate quantum angular momenta specifying the conformal block 
representation. Each one of these quantum numbers belongs to the finite set $\left\{ 0,\frac{1}{2}, ... ,\frac{k}{2} 
\right\}$ (due to the finite number of allowed primary fields for the relative conformal field theory) and therefore  
a register of $\lceil \log \left( k + 1 \right) \rceil$ qubits is necessary for encoding for each one of them. The 
total number of qubits needed to encode for a basis vector is the product of the number of observables that specify 
a state times the number of qubits needed to specify a quantum number: $\left( 4m-3\right) \lceil \log \left( k + 1 
 \right)\rceil$, where $2m$ is the index of the braid group. This quantity grows linearly in the index of the braid group. An ordering on the quantum register must also be 
chosen, such that it is possible to associate to each block of $ \lceil \log \left( k + 1 \right) \rceil $ qubits a 
precise quantum number of the system. The ordering we choose is the one shown in figure [\ref{register}].

\noindent The action of a unitary matrix on the encoding space $E$ is analogue to the action of the same matrix 
on the physical space in the following sense. If $$U : | p \rangle \mapsto | q \rangle \; ,$$ then $$E \left( U 
\right) : | E \left( p \right) \rangle \mapsto | E \left( q \right) \rangle \; .$$ The encoding space constitutes 
a new representation space for the quantum system.

\begin{figure}
\begin{center}
\includegraphics[height=6cm]{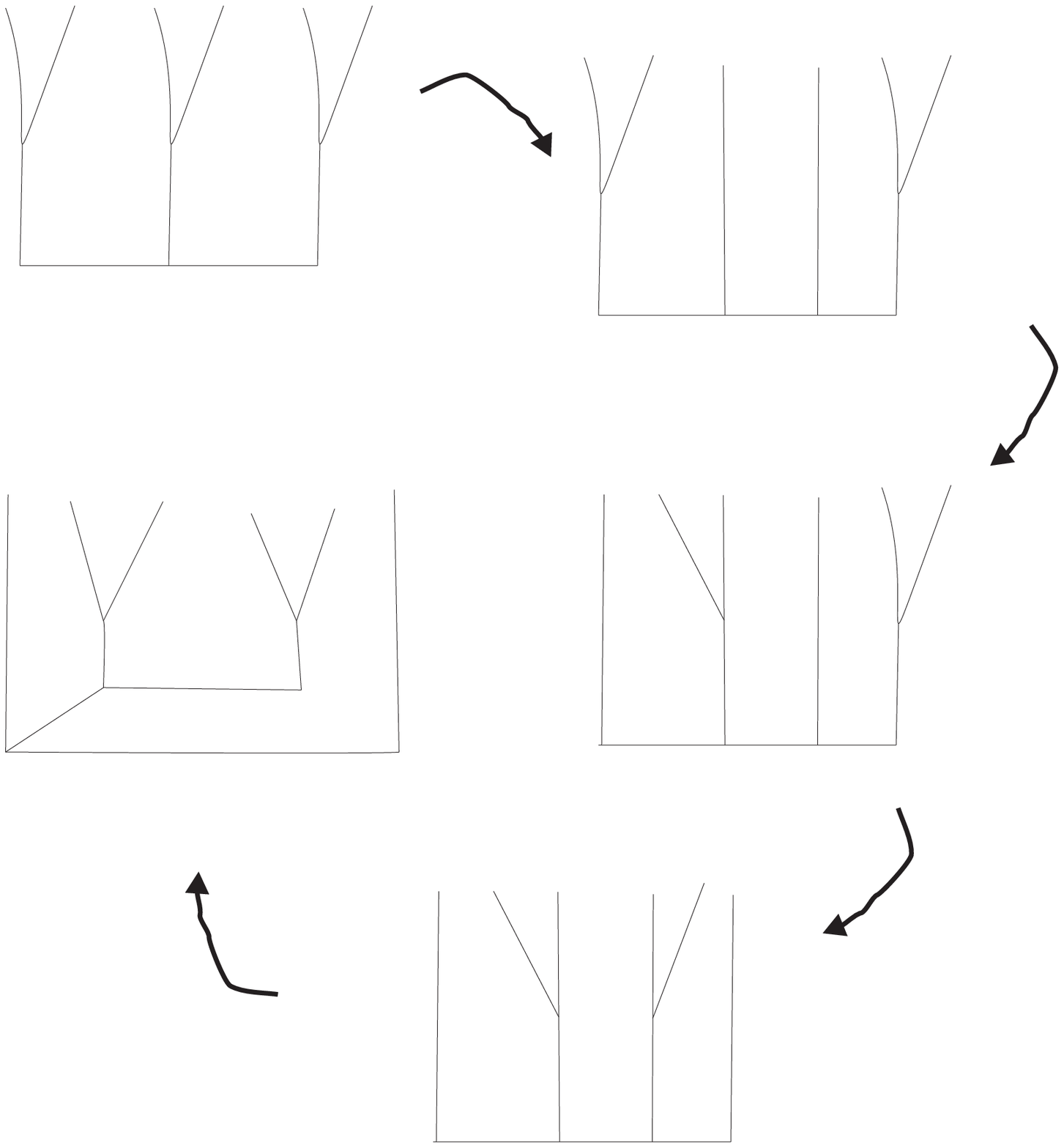}
\caption{\footnotesize{Duality decomposition of 6-point correlator.}}                                           \end{center}
\label{6point}
\end{figure}

\noindent Using the Kaul representation for the braid group, even-indexed elements of $B_{2m}$ are mapped to 
unitary matrices which are diagonal in the representation space, while odd-indexed elements of $B_{2m}$ are mapped 
to unitary matrices constructed with duality matrices. Each duality matrix, in turn, can be decomposed in a sequence 
of elementary duality matrices, that correspond to $q-6j$ recoupling transformations, using the following relation 
\begin{eqnarray}
&\,& A_{\left( \textbf{p};\textbf{r}\right) }^{\left( \textbf{q};\textbf{s}\right) }\left [ 
\begin{tabular}{cc}
 $j_1$ & $j_2$ \\
 $\vdots$ & $\vdots$ \\
 $j_{2m-1}$ & $j_{2m}$ \\
\end{tabular} 
\right ] 
= \\ \nonumber 
&\,& = \sum_{t_1,t_2,...,t_{m-2}} \prod_{i=1}^{m-2} \left( A_{p_i}^{t_i} \left [ 
\begin{tabular}{cc}
$r_{i-1}$ & $j_{2i+1}$ \\
$j_{2i+2}$ & $r_i$ \\
\end{tabular} 
\right ] A_{t_i}^{s_{i-1}}
\left [ 
\begin{tabular}{cc}
$t_{i-1}$ & $q_{i}$ \\
$s_{i}$ & $j_{2m}$ \\
\end{tabular} 
\right ] 
\right ) \times \\ \nonumber 
&\,& \times \prod_{l=0}^{m-2} A_{r_l}^{q_{l+1}}
\left [ 
\begin{tabular}{cc}
$t_{l}$ & $j_{2l+2}$ \\
$j_{2l+3}$ & $t_{l+1}$ \\
\end{tabular} 
\right ] 
\end{eqnarray} 
The above decomposition allows us to build a quantum circuit that results to be crucial for the algorithm.  
In order to have a relatively simple example of graphical realization of the decomposition one can compare  
fig.[\ref{6point}] and the corresponding quantum circuit in fig.[4].

\begin{figure}
\begin{center}
\includegraphics[height=6cm]{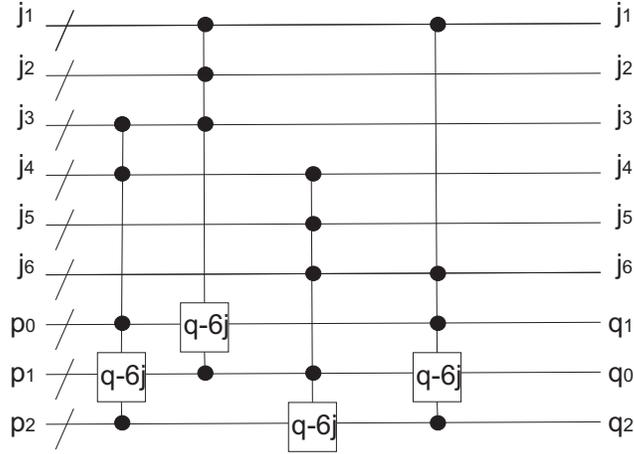}
\caption{\footnotesize{Quantum circuit for the 6-point quantum correlator.}}
\end{center}
\label{circuit}
\end{figure}
 
The above decomposition corresponds to a path on the $q$-deformed spin network simulator. Each element of a $q-6j$ 
gate can be efficiently evaluated by a classical machine. Since a $q-6j$ gate acts always on a $5 \lceil \log (k+1)  
\rceil$-qubits register, it can be efficiently compiled with a set of universal matrices \cite{HRC}. 
Increasing the size of the input will result in an increasing number of $q-6j$ gates in the decomposition of the 
$q-3nj$ gate, but each $q-6j$ gate will always be $2^{5 \lceil \log (k+1) \rceil}$-dimensional. A careful examination of 
the graphical realization of the above decomposition allows then us to estimate the increase rate, with respect to the 
input size, of the number of $q-6j$ gates. The dependence on $2m$, index of the braid group, of $g$, 
number of $q-6j$ gates, is given by $g=3m-5$. 

The above considerations prove that it is possible to efficiently compile Kaul's unitary representation of the 
colored braid groupoid. In order to prove the efficiency of the quantum algorithm for the approximation of the 
colored Jones polynomial we shall resort now to the standard procedure to access information on a quantum system 
by a series of measurements. 

\subsection{Quantum algorithm for knot invariants.} 
In this subsection we give a polynomial-time quantum algorithm which approximates the value of the colored Jones 
polynomial of a link obtained as the plat closure of a braid. The same algorithm admits also a realization 
which holds for links obtained as standard closure of braids. The notion of approximation used in this context 
was formalized by \cite{BFL}. Adopting the terminology of \cite{WY} by \textbf{additive approximation} we mean the 
following: given a normalized function $f(x)$, where $x$ denotes an instance in the selected coding, we have an 
additive approximation of its value for each instance $x$ if we can associate to it a random variable $Z_x$ such 
that 
$$
{\rm Pr} \left\{ \left | f(x) - Z_x \right | \leq \delta \right \} \geq 3/4 \; , 
$$ 
for any $\delta \geq 0$. The time needed to achieve the approximation must be polynomial in the size of the problem 
and in ${\delta}^{-1}$. The normalization adopted for the colored Jones polynomial of a link is provided by the 
product, over all the component knots, of the quantum dimension of the knots. By \textit{coloring} we mean the set of possibly different irreps of $SU(2)_q$ labeling the component knots of a link. The problem we are interested in can 
now be stated as follow.  

\smallskip 

\noindent \textbf{Problem: Approximate colored Jones.} Given a colored braid $b \in B_{2m}$ of length $\ell$, a 
coloring $\textbf{c}$, a positive integer $k$ and $\delta > 0$, we want to sample from a random variable $Z$ which 
is an additive approximation of the absolute value of the colored Jones polynomial of the plat closure, evaluated 
at $q = \exp \left ( {\frac{2 \pi i}{k+2}}\right )$, such that the following condition holds true \[ {\rm Pr} \left ( 
\left | V (L, \textbf{c}, q) - Z \right | \leq \delta \right ) \geq 3/4 \;. \] In the following we hall provide an 
efficient quantum algorithm to \textbf{approximate colored Jones}, which solves the problem in $O ({\rm poly} (\ell , 
{\delta}^{-1}))$ steps.  

\noindent As in \cite{AJL,WY} we need the following two lemmas in order to prove the efficiency of the algorithm.

\smallskip 

\noindent \textbf{Lemma 1.} Given a quantum circuit $U$ of length $O ({\rm poly} (n))$, acting on $n$ qubits, and 
given a pure state $| \psi \rangle$ which can be prepared in time $O ({\rm poly} (n))$, then it is possible to sample  
in $O ({\rm poly} (n))$ time from two random variables $a$ and $b$, which can assume only the values +1 or -1, in such 
a way  that $< a + ib > = \langle \psi | U | \psi \rangle$.

\smallskip 

\noindent \textbf{Lemma 2.} Given a set of random variables $\{ r_i  \}$ such that
$$
r_i \geq 0 \quad ,\quad {n}^{-1} \, {\sum r_i} \stackrel{n \rightarrow \infty}{\longrightarrow} m \quad ,\quad  
{n}^{-1}\, {\sum \bigl ( r_i^2 - \langle r_i^2 \rangle} \bigr ) \stackrel{n \rightarrow \infty}{\longrightarrow} 
v \; , 
$$ 
then ${\rm Pr} \left( \left | {n}^{-1}\, \sum r_i - m \right | \geq \delta \right ) \leq 2\, \exp \bigl ( - 
n \delta^2 / (4 v) \bigr).$

\smallskip

\noindent The first lemma can be easily proved with linear algebra arguments \cite{AJL,WY}. The second lemma is a modified 
version of the well known Chernoff bound.

\begin{figure}
\begin{center}
 \includegraphics[height=3cm]{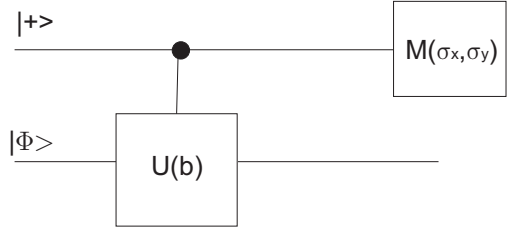}
\end{center}
\caption{\footnotesize{Quantum circuit for the evaluation of the colored Jones polynomial, where $| + \rangle = \frac{1}{2} \left(|0\rangle + |1\rangle \right)$ and $|\Phi \rangle = | \Phi_{\left( \textbf{0};\textbf{0} \right)} \rangle$}.}
\label{algo}
\end{figure}

\noindent We can conclude now that the qubit model for the Kaul representation can be used to efficiently compile 
a unitary representation of the colored braid group, and a sampling procedure can then be used to efficiently access 
the value of the colored Jones polynomial. The circuit that realizes all these steps is schematized in fig.[\ref{algo}]. 
The sampling lemma tells us that a series of measurements of $\sigma_{x}$ on the first qubit will provide the value 
for ${\rm Re}\, \left ( V \left ( L,\textbf{c}, q \right )\right )$, while a series of measurements of $\sigma_{y}$ on the 
first qubit will provide the value for ${\rm Im}\, \left ( V \left ( L,\textbf{c},q \right )\right )$.

\section{Conclusions.}
We have provided a quantum algorithm that efficiently approximates the colored Jones polynomial. The construction 
is based on the exact solution of Chern-Simons TQFT and its connection to WZW conformal filed theory. We showed 
elsewhere \cite{GaMaRa1} that the colored Jones polynomial can be seen as the expectation value of the evolution of the 
$q$-deformed spin-network quantum automaton. A quantum circuit capable of evaluating this probability amplitude, 
given as input a braid word, will also approximate the evolution of the automaton. Since the automaton processes  
efficiently each braid word, if the quantum circuit efficiently simulates the automaton then it will also provide 
an efficient approximation of the colored Jones polynomial. The evaluation of the invariant is equivalent to the 
evaluation of the expectation value for some unitary operators. It has been proved that using the Hadamard trick 
it is possible to efficiently approximate this expectation value, once an efficient decomposition of the operator 
has been found and if the initial state of the system is efficiently obtainable. The present work satisfies all 
the above requirements. A more detailed exposition will be given in forthcoming  papers.

\end{document}